\documentclass[conference]{IEEEtran}
\usepackage{amsmath,amssymb}

\usepackage{comment}
\usepackage{amssymb}
\usepackage{bm}
\usepackage[dvipdfmx]{graphicx}
\usepackage[dvipdfmx]{color}
\graphicspath{{figure/}}
\usepackage{url}

\usepackage{setspace}

\begin{document}
\title{Sparse Signal Recovery for Binary Compressed Sensing  by Majority Voting Neural Networks} 

\author{
  \IEEEauthorblockN{Daisuke Ito and Tadashi Wadayama}
  \IEEEauthorblockA{Department Computer Science and Engineering, \\Nagoya Institute of Technology,
    Nagoya,  Japan\\
    Email: d.ito.480@stn.nitech.co.jp, wadayama@nitech.ac.jp} 
}


\graphicspath{{figure/}}

\maketitle
\begin{abstract}
In this paper, we propose majority voting neural networks 
for sparse signal recovery in binary compressed sensing.
The majority voting neural network is composed of
several independently trained feedforward neural networks 
employing the sigmoid function as an activation function.
Our empirical study shows that a choice of a loss function used in training processes for the network is 
of prime importance. We found a loss function 
suitable for sparse signal recovery, which includes a cross entropy-like term and an $L_1$ regularized term. 
From the experimental results, we observed that  the majority voting neural network
achieves excellent recovery performance, which is approaching the optimal performance 
as the number of component nets grows.
The simple architecture of the majority voting neural networks would be beneficial for 
both software and hardware implementations.
\end{abstract}

\section{Introduction}

Recent flourish of studies on compressed sensing inspires researchers 
to apply the idea for applications in wireless communications.
Fletcher et al. \cite{onoff} showed that
a detection problem for on-off random access channels  
is equivalent a sparse signal recovery problem discussed in compressed sensing.
Kaneko et al. \cite{RFID} used a compressed sensing technique to develop 
an identification protocol for a passive RFID system.
For such an application in wireless communications, we need powerful sparse signal recovery algorithms
that are suitable for hardware implementation in order to achieve high speed signal processing 
and energy efficiency.

{\em Binary (or one-bit) compressed sensing}
firstly proposed by Boufounos and Baraniuk \cite{1bit_comp} is a variant of the original compressed sensing. 
In the scenario of binary compressed sensing,  a linear observation vector 
that is quantized to a binary value, typically in the binary alphabet $\{+1, -1\}$,  is given to a receiver.
The linear observation vector is the product of a sensing matrix and a hidden sparse signal vector.
A receiver then tries to reconstruct the original sparse signal. This process is called a 
{\em sparse signal recovery} process.
It is known that power consumption of an AD converter is closely related to 
the number of its quantization levels and the sampling frequency,
namely, the power consumption of an AD converter 
increases as the sampling frequency increases. 
When we pursue to develop an extremely low power consumption device for 
battery-powered sensors  or to develop a digital signal processing device 
operating at very high sampling frequency,  binary quantization by a comparator is a reasonable choice. 
Binary compressed sensing is well suited to such a situation. 
Moreover, since the input to the receiver is restricted to binary values,  
no gain control is required in the case of binary compressed sensing.
This fact further simplifies the hardware  needed  in the receiver.

The optimal sparse signal recovery for binary compressed sensing 
can be attained by solving a certain integer programming problem. 
However, the IP problem is, in general,  computationally hard to solve 
and the approach can handle only small problems. 
Boufounos and Baraniuk \cite{1bit_comp} studied a relaxation method that replaces 
$L_0$-norm with $L_1$-norm and introduced a convex relaxation for integer constraints.
Although nonlinearity induced by  binary quantization prohibits
direct applications of known sparse recovery algorithms for the original compressed sensing,
several sparse recovery algorithms for binary compressed sensing 
has been developed \cite{greedy} \cite{BIHT} \cite{RSS} 
based on the known iterative methods for the original compressed sensing.
Boufounos \cite{greedy} proposed a greedy algorithm called Matched Sign Pursuit (MSP) that is 
a counter part of Orthogonal  Matching Pursuit (OMP). 
The paper \cite{BIHT} presents  Binary Iterative Hard Thresholding (BIHT) algorithm 
by reforming Iterative Hard Thresholding (IHT) algorithm \cite{IHT}.

Although the known sparse recovery algorithms exhibit reasonable sparse recovery performance, 
it may not be suitable for applications in high speed wireless communications. This is because 
most of algorithms require a number of iterations to achieve reasonable sparse recovery results.
Most of known algorithms also require calculations involving
matrix-vector products with $O(n^2)$-computations for each iteration where 
$n$ is the length of the sparse signal vector.

Our approach for sparse signal recovery is to employ {\em feedforward neural networks} as 
building blocks of sparse signal recovery schemes.
We expect that an appropriately designed and trained neural networks 
greatly reduce required computing resources and are well suited to hardware implementation.
In this paper,  we will propose the majority voting neural networks composed of several 
independently trained neural networks, 
which are feedforward 3-layer neural networks 
employing the sigmoid function as an activation function.
As far as we know, there is no previous studies on sparse signal recovery based on neural networks.
Our focus is thus not only on the practical aspect of the neural sparse signal recovery but also 
on studies on the fundamental behavior of the neural networks for sparse signal recovery. 
Recently,  {\em deep neural networks} \cite{deep_net} have been actively studied 
because they provide surprisingly excellent performance in the areas of 
image/speech recognition and natural language processing \cite{very_deep}. 
Such powerful neural networks can be used in wireless communication as well.
This work can be seen as a first attempt to this direction.

\section{Sparse Recovery by Neural Networks}\label{NN_construction}

\subsection{Binary compressed sensing}\label{subsec:comp}

The main problem for binary  compressed sensing is to 
reconstruct an unknown sparse signal vector $x \in \Bbb R^n$ from the observation 
signal vector $u \in \{+1, -1\}^m$
under the condition that these signals satisfy the relationship:
\begin{equation}
u = {\rm sign}(Ax).
\end{equation}
The  sign function ${\rm sign}(\cdot)$ is defined by
\begin{equation}
{\rm sign}(a) = \begin{cases}
-1 & (a \le 0),\\
+1 & (a > 0).
\end{cases}
\end{equation}
The matrix $A \in \Bbb R^{m \times n}$ is called  a {\em sensing matrix}.
We assume that the length of the observation signal vector $u$ is smaller than the 
length of the sparse signal vector $x$, i.e., $m < n$.
This problem setup is similar to that  of the original compressed sensing.
The notable difference between them is that the observation signal $u$
is binarized in a sensing process of binary compressed sensing.
A receiver obtains the observation signal $u$ and then it 
tries to recover the corresponding hidden signal $x$.
We here make two assumptions for the signal $x$ and the sensing matrix $A$.
The first assumption is sparsity of the hidden signal $x$.
The original binary signal $x \in \Bbb \{0,1 \}^n$ contains only $k$ non-zero elements,
where  $k$ is a positive integer much smaller than $n$,
i.e., Hamming weight of $x$ should be $k$.
We call the set of binary vectors with Hamming weight $k$ is $k$-{\em sparse signals}.
The second assumption is that the receiver completely knows the sensing matrix $A$.

\subsection{Network architecture}

When we need an extremely high speed signal processing or 
an energy-efficient sparse signal processing method for battery powered sensor,
it would be reasonable to develop a sparse signal recovery algorithm suitable 
for the situation. In the sparse signal recovery method based on neural networks to be described in this section
requires only several matrix-vector products to obtain an output signal, 
which is an estimate signal of the sparse vector $x$.
Thus, the proposed method needs smaller computational costs than those
required by conventional iterative methods.

\begin{figure}[tbp]
        \begin{center}
          \includegraphics[clip, width=7cm]{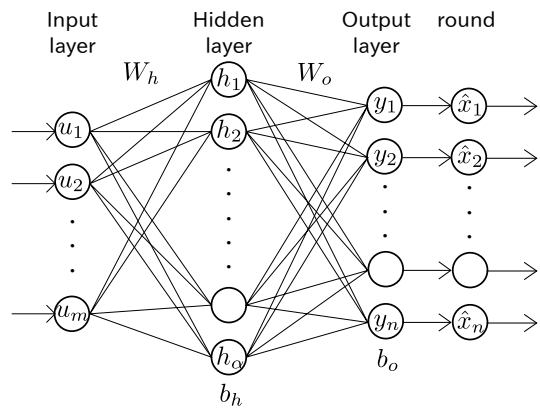}
          \caption{Architecture of feedforward neural networks for sparse signal recovery. An observation signal $u$ 
          comes from the left and is fed to the input layer. A sparse estimation vector comes out from the output layer.  
          The sigmoid function is used as an activation function.}
          \label{fig:NNmodel}
        \end{center}
\end{figure}

Our sparse recovery method is based on a 3-layer feedforward neural network illustrated in Fig.\ref{fig:NNmodel}.
This architecture is fairly common one; it consists of the input, hidden and output layers.
Adjacent layers are connected by weighted edges and each layer includes neural units that can keep real values.
As  an {\em activation function}, we employed the sigmoid function to determine the values of the hidden and output
layers. In our problem setting,  the observation signal $u$ is fed into the input layer from the left in Fig.\ref{fig:NNmodel}.
The signal propagates from left to right and the output signal $y$ eventually comes out from the output layer.
The network should be trained so that the output signal $y$ is an accurate estimation of the original sparse 
signal $x$.
The precise description of the network in Fig.\ref{fig:NNmodel} 
is given by the following equations:
\begin{align}
h = \sigma(W_h u + b_h), \label{hyde}\\
y = \sigma(W_o h + b_o), \label{output} \\
\hat x = \mbox{round}(y). \label{round}
\end{align}
The function $\sigma(\cdot)$ is the sigmoid function defined by
$
f(a) = {1}/{(1 + e^{-a})}.
$
In this paper, we will follow a simple convention that
$f(a)$ represents $(f(a_1), f(a_2), \ldots, f(a_n))$ where $a = (a_1,\ldots, a_n)$.
The round function ${\rm round}(a) (a \in  \Bbb R)$ gives the nearest integer from $a$.
The equation (\ref{hyde}) defines the signal transformation from the input layer to the hidden layer.
An affine transformation is firstly applied to the input signal $u \in \{+1, -1\}^n$ and then 
the sigmoid function is applied. The weight matrix $W_h \in \mathbb{R}^{m \times \alpha}$ and 
the bias vector $b_h \in \mathbb{R}^{\alpha}$ defines the affine transformation.
The resulting signal $h \in \Bbb R^{\alpha}$ is kept in the units in the hidden layer, which are called 
{\em hidden units}.
The parameter $\alpha$ thus means the number of hidden units.
From the hidden layer to the output layer,  the equation (\ref{output}) governs the second signal 
transformation. The second transformation to yield the signal $y$
consists of the affine transformation,  based on the weight matrix $W_o \in \mathbb{R}^{\alpha \times n}$
and the bias vector $b_o \in \mathbb{R}^n$, and the nonlinear mapping based on the sigmoid function.
The vector $y \in {\Bbb R}^n$ emitted from the output layer is finally rounded to a nearest integer vector because 
we assumed that non-zero elements in the original sparse signal $x$ takes the value one.
Since the range of the sigmoid function lies in the open interval $(0, 1)$, 
an element in the estimate vector $\hat x$ 
should take the value zero or one.

\subsection{Training}\label{subsec:NN_study}

The network in Fig.\ref{fig:NNmodel} can be seen as a parametrized estimator
$
\hat x = {\rm round}(\Phi_\theta (u))
$
where $\theta$ is the set of the trainable parameters $\theta = \{W_h,W_o,b_h,b_o\}$.
It is expected that the trainable parameter $\theta$ should be adjusted in the training phase 
so as to minimize the error probability $Prob[x \ne \hat{x}]$. 
However, it may be computationally intractable to minimize the error probability directly.
Instated of
direct minimization of the error probability itself,  we will minimize a loss function 
including a cross entropy-like loss function and an $L_1$-regularization term.
In this subsection, the details of the training process  is described.

In the training phase of the network,
the parameter $\theta$ should be updated in order to minimize the values of the given loss function.
Let $
\mathcal{D} = \{ (u^1,x^1), \ldots, (u^L, x^L) \}
$
be the set of the training data used in the training phase.
The signals $u^i$ and $x^i$ relate as $u^{i} = {\rm sign}(Ax^i), i = 1, 2, \ldots, L$.
In the training process, we use randomly generated training samples;
the sparse vectors $x^1, x^2, \ldots, x^L$ are generated uniformly at random from the set of $k$-sparse vectors.
The sample $u^i$ is fed into the network and 
the corresponding sample $x^i$ is used as the {\em supervisory signal}.

We here employ {\em stochastic gradient descent (SGD) algorithms} to 
minimize the loss function described later.
It is empirically known that SGD and its variations behave very well for 
non-convex objective functions that are computationally hard to minimize. 
This is the reason why SGD and related algorithms are widely used for 
training deep neural networks. 
In order to use SGD, we need to partition the training set into minibatches.
A minibatch is a subset of the training data and the use of minibatches 
introduces stochastic disturbance in training processes.
Such stochastic disturbance helps 
a search point in an SGD process to escape from a stationary point of 
the non-convex objective function to be minimized. 
We divide the training data into a number of minibatches as follows:
\begin{equation}
\mathcal{D} = \{\{(u^{1},x^{1}), \ldots, (u^{T},x^{T})\},\{(u^{T+1},x^{T+1})\ldots, 
\}\ldots\}. \nonumber
\end{equation}
In this case, every minibatch contains $T$-pair of samples.
We denote  $k$-th $, k= 1,2,\ldots,L/T$,  minibatch as $\mathcal{D}_k$.

\subsection{Loss function}

The choice of the loss function, i.e., the objective function to be minimized in training processes, 
is crucial to achieve appropriate recovery performance.
We  introduce a loss function designed for sparse signal recovery. 
The loss function of $k$-th minibatch is defined by
\begin{equation}\label{loss_k}
L_k(\theta) = -\frac{1}{n T}\sum_{i \in \mathcal{D}_k}  \sum_{j=1}^n x^i_j \log y^i_j + \frac{\lambda}{T} \sum_{i \in \mathcal{D}_k} ||y^i||_1.
\end{equation}
The vector $y^i = (y^i_1,\ldots, y^i_n)$ is given by
$
y^i = \Phi_\theta (u^i),
$
i.e, the output of our neural network corresponding to the input $u^i$.
The first term of $L_k(\theta)$ measures closeness between $y^i$ and $x^{i}$.
This measure is closely related to {\em cross entropy} that are often used in supervised classification problems. 
In the case of a classification problem,  $x^i$ is a one-hot vector that can be interpreted as a probability vector. 
In our case, since $x^{i}$ contains $k$-ones,  thus the first term  is not the same as the cross entropy.
It has been empirically observed that this term plays an important role for sparse signal recovery. For example, 
from several numerical experiments indicated that the $L_2$-distance between $y^i$ and $x^{i}$ is not suitable 
for our purpose.
The second term of equation (\ref{loss_k}) is the $L_1$-{\em regularization term} for promoting sparsity of the output $y$.
The regularization parameter $\lambda$ adjusts effectiveness of regularization.
Some experiments showed that the $L_1$-regularization term is indispensable for obtaining sparse output vector.

The training process of our network can be summarized as follows.
We first generate the training data $\mathcal{D}$. 
In the $k$-th update iteration of the parameter $\theta$,  the minibatch $\mathcal{D}_k$ is fed to 
the Adam optimizer \cite{adam} based on the loss function (\ref{loss_k}). 
The Adam optimizer is a variant of SGD that provides fast convergence in many cases 
and it is widely used in learning process for deep neural networks.
An iteration corresponding to a process for a minibatch is called a {\em learning step}.
A training process finishes when all the minibatches are processed.

\section{Numerical Results}

As the primal performance measure of sparse signal reconstruction, 
we adopt the {\em recovery rate} which is the probability of the 
event $\hat x \ne x$ under the assumption where $x$ is chosen uniformly 
at random from the set of $k$-sparse binary vectors.
In this section, we evaluate the sparse signal recovery performance of 
our feedforward neural networks in Fig. \ref{fig:NNmodel}.

\subsection{Details on experiments}

We used TensorFlow \cite{tens} to implement and to train our neural networks.
TensorFlow is a framework designed for distributed data flow based numerical calculations 
that is especially well suited for training of deep neural networks.
TensorFlow supports automatic back propagation for computing the gradient vectors
required for the parameter updates and it also provides 
GPU-computing that can significantly accelerate training processes. 
It is rather straightforward to implement 
our neural networks and the training process descried in the previous section by using TensorFlow.

The details of parameters used throughout the paper are following.
The length of the original sparse signal $x$ is set to $n = 256$.
The sensing matrix $A \in \Bbb R^{m \times n}$ is generated at random, i.e., 
each element of $A$ is independently generated according to Gaussian distribution with mean $0$ and variance $1$.
The sensing matrix is generated just before an experiment and fixed during an experiment.
At the beginning of a training process, 
the weight matrices $W_h$ and $W_o$ at the hidden and output layers
are initialized based on pseudo random numbers, namely each element of these matrices 
follows Gaussian distribution with mean $0$ and variance $0.05$.
The bias vectors $b_h$ and $b_o$ at the hidden and output layers
are initialized to the zero vector.
The number of hidden units is set to $\alpha=1000$ and
the minibatch size is set to $T=100$.
The initial value of the learning coefficient of Adam optimizer is $\epsilon = 0.002$ and
the coefficient of regularization term is set to $\lambda = 0.95$.

\subsection{Initial experiments}

\begin{figure}[tbp]
        \begin{center}
          \includegraphics[width=8.0cm]{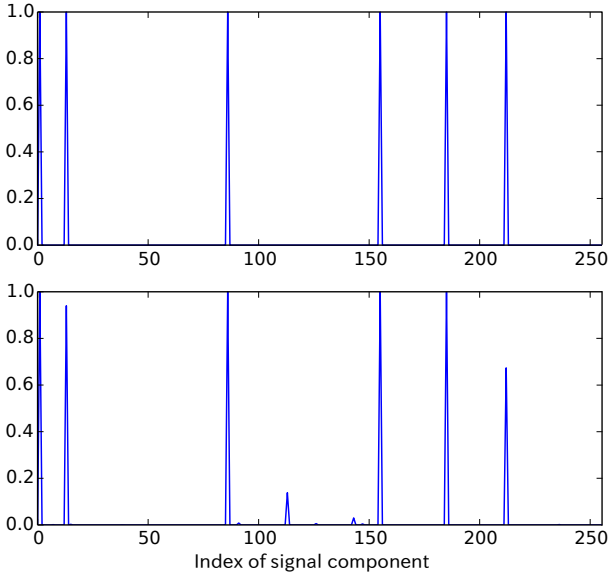}
          \caption{Sparse signal recovery for a $6$-sparse vector.
          (top: the original sparse signal  $x$,  bottom: the output $y = \Phi_{\theta^*}(x)$ from the trained neural network.
           $n=256, m=120$)
          }
          \label{fig:NN_output}
        \end{center}
\end{figure}

As an example of sparse signal reconstruction via our neural network, 
we show a $6$-sparse vector $x$ and 
the corresponding output of the trained neural network $y = \Phi_{\theta^*}(x)$ in Fig.\ref{fig:NN_output}.
The parameter $\theta^*$ is obtained by a training process 
with $5 \times 10^4$ learning steps, i.e., minibatches. 
In this experiment, we set the length of the observation signal to $m=140$.
From Fig.\ref{fig:NN_output}, 
we can observe that the output $y$ shows fairly good match with the original signal $x$.
For example, the support of components in $y$ with the value larger than $0.5$
exactly coincides with the support of $x$. Some of the components have the value pretty close to $1$ as well.
It is also seen that some of components with small values, e.g., at indices around 110 and 145,  incur
false positive which means that the corresponding components in the original $x$ are zero.
This is the reason why we introduced the round function at the final stage of our neural network in (\ref{round}).
It is expected that the round function eliminates the effect of the components with small values that may
produce false positive elements in the final estimation $\hat x$.

\begin{figure}[tbp]
        \begin{center}
          \includegraphics[clip, width=8.4cm]{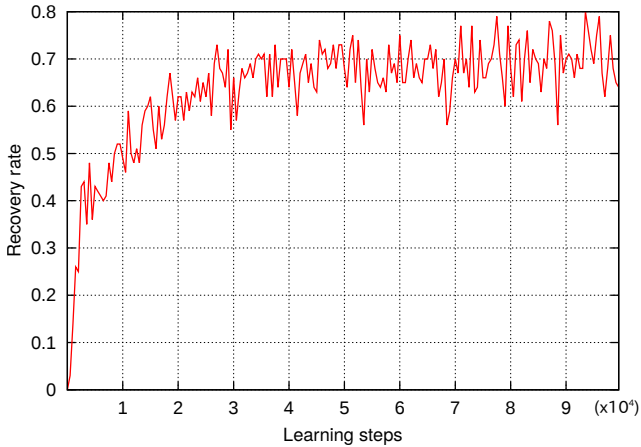}
          \caption{Learning steps and recovery rate.
          $(n=256,  m=140, k=6)$}
          \label{fig:RR_steps}
        \end{center}
\end{figure}

Fig.\ref{fig:RR_steps} indicates a relationship between the recovery rate and the learning steps.
The parameters are $n=256,  m=140, k=6$.
From Fig.\ref{fig:RR_steps}, we can see that 
the recovery rate increases as the number of  learning steps increases 
although the progress contains fluctuations.
The recovery rate appears to be saturated around $5.0 \times 10^4$ steps.
In following experiments, the number of learning steps is thus set to $5 \times 10^4$ based on this observation.

\subsection{Sparse recovery by integer programming}\label{sec:IP}

We introduce {\em integer programming (IP)-based sparse signal recovery}  as  a performance benchmark
in the subsequent subsections because it provides the optimal recovery rate.
The IP formulation shown here is based on the linear programming formulation in \cite{IP}. 
Although IP-based sparse signal recovery requires huge computer resources, 
it is applicable to moderate size problems if we employ a recent advanced IP solver.
We used IBM CPLEX Optimizer  for solving the IP problem shown below.
The problem needed to solve is 
to find a feasible binary vector $z = (z_1, \ldots, z_n)  \in \{0,1\}^n$ satisfying the following conditions:
\begin{eqnarray}
&&  \sum_{i=1}^n z_i = k, \\
&& \forall i \in [1,m],\quad \sum_{j = 1}^n A_{i,j} z_j > 0,\quad \mbox{if } u_i = +1, \\
&& \forall i \in [1,m],\quad \sum_{j = 1}^n A_{i,j} z_j \le 0,\quad \mbox{if } u_i = -1, 
\end{eqnarray}
where $A_{i,j}$ is the $(i,j)$ element of the sensing matrix $A$ and $u_i$ is the $i$-th element 
of the observation signal $u$. If a feasible solution $z^*$ satisfying all the above conditions exists, 
it becomes an estimate $\hat x = z^*$.
It is clear that these conditions are consistent with our setting of 
binary compressed sensing.

\subsection{Experimental results}\label{subsec:NN}

\begin{figure}[tbp]
        \begin{center}
          \includegraphics[scale=1.0]{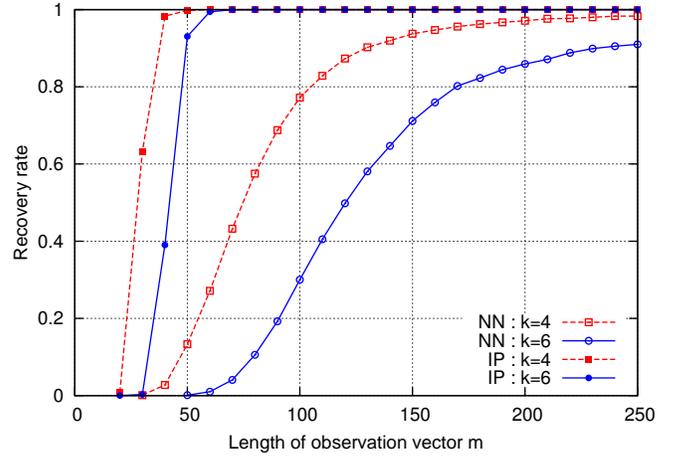}
          \caption{Recovery rates of the proposed scheme (denoted by {\bf NN}). ($n=256, k=4, 6$) As benchmarks, 
          recovery rates of IP-based scheme is also included (denoted by {\bf IP}).
            }
          \label{fig:RR_normal}
        \end{center}
\end{figure}

Fig.\ref{fig:RR_normal} presents the recovery rates of the proposed scheme 
under the condition where $n = 256, k=4, 6$.
The neural network shown in Fig.\ref{fig:NNmodel} was used.
In Fig.\ref{fig:RR_normal},
it is seen that the recovery rate tend to increase as  $m$ increases for all $k$.
The recovery rate is beyond  $90\%$ at $m=140$ when the original signal is $4$-sparse.
It can be also observed that the recovery rate strongly depends on the sparseness parameter $k$.
For example, the recovery rate of $k=6$ comes around $70\%$ at $m=140$.
The IP-based sparse recovery provides  
the recovery rate  $99\%$ at $m \ge 60$ when the original signal is $6$-sparse vector.
On the other hand, our neural network yields the recovery rate more than $90\%$ at $m \ge 240$ 
when the original signal is $6$-sparse.
Although computation costs of the neural network in the recovery phase are much smaller than 
those required for IP-based sparse recovery,  the performance gap appears rather huge and 
the neural-based reconstruction should be further improved.

\section{Majority Voting Neural Networks}

In the previous section, we saw that neural-based sparse signal recovery is 
successful under some parameter setting but there are still much room for improvement 
in terms of the recovery performance. 
In this section, we will propose a promising variant of the feedforward neural networks
which is called {\em majority voting neural networks}.
The majority voting neural network consists of several independently trained neural networks.
The outputs from these neural network are combined by  soft majority voting nodes and 
the final estimation vector is obtained by rounding the output from the soft majority voting nodes.
Combining a several neural networks to obtain improved performance
is not a novel idea, e.g., \cite{majority},   but it will be shown that the idea is very effective for 
our purpose.

\subsection{Network architecture}

From statistics of reconstruction errors occurred in our computer experiments, we observed that 
many reconstruction error events (i.e., $x \ne \hat x$) occur due to only one symbol mismatch.
In addition to this observation,
we also found that independently trained neural networks tend to make symbol errors 
at distict positions.
These observations inspire us to use majority voting to combine several outputs from 
independently trained neural networks.

\begin{figure}[tbp]
        \begin{center}
             \includegraphics[clip, width=8.4cm]{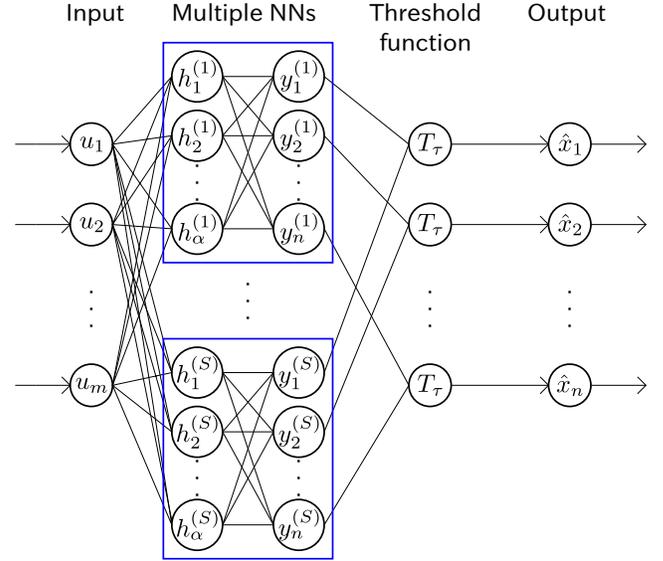}
          \caption{Network architecture of majority voting neural networks. The netowork consists of 
          $S$ independently trained feedforward neural networks.
}
          \label{fig: MVNN}
        \end{center}
\end{figure}

Figure \ref{fig: MVNN} presents the architecture of the majority voting neural networks.
In this case, the majority voting neural network consists of $S$ component feedforward neural networks
defined by
\begin{align}
h^{(s)} = \sigma(W_h^{(s)} u + b_h^{(s)}),  \\
y^{(s)} = \sigma(W_o^{(s)} h^{(s)} + b_o^{(s)}),  
\end{align}
where $s = 1,2,3, \ldots,S$. The output of the component neural networks are aggregated by the soft
majority logic nodes and  it yields the estimation vector:
\begin{equation}
\hat x = T_\tau \left(\sum_{s=1}^S y^{(s)} \right),
\end{equation}
where $T_\tau(\cdot)$ is the the threshold function defined by
\begin{equation}
T_\tau(a) =
\left\{
\begin{array}{cc}
0, &  \mbox{if } a \le \tau, \\
1, &  \mbox{if } a > \tau.
\end{array}
\right.
\end{equation}
In the following experiments, we set $\tau = S/2$.
Each component network was trained independently. This means that the training sets were
independently generated for each component network.  

\subsection{Required computational resources}

The simple architecture of the majority voting neural networks is advantageous 
for both software and hardware implementations. In a case of software implementation,
computation time required for matrix-vector products are dominant in a recovery process.
For computing $y^{(s)}$, we need approximately $2 \alpha (m+n)$  basic arithmetic operations such as  
additions and multiplications. 
Since there are $S$-components,  approximately $2 S \alpha (m+n)$  basic arithmetic operations
are required for computing the output.
This number appears competitive to known iterative methods \cite{greedy} because, 
in most iterative algorithms,  $O(n^2)$-basic operations are required for each iteration to compute $A x'$ where 
$A \in \Bbb R^{m \times n}$ and $x' \in \Bbb R^n$.
For a case of hardware implementation,  {\em parallelism in the architecture} of the majority voting neural networks
possibly enables us to create high speed sparse recovery circuits on FPGA or ASIC.
Note that implementation of neural networks with FPGA is recently becoming a hot research topic \cite{FPGA}.

\subsection{Experimental results}

\begin{figure}[tbp]
        \begin{center}
          \includegraphics[clip, width=8.4cm]{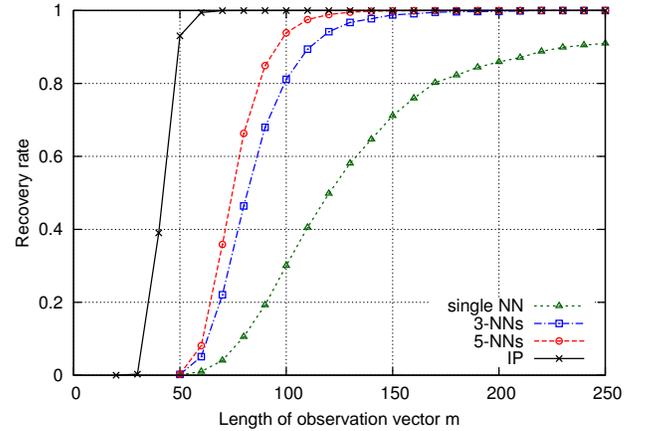}
          \caption{Recovery rate of majority voting neural network.  ($n=256, k=6$)}
          \label{fig:RR_and_m_k6}
        \end{center}
\end{figure}

Fig.\ref{fig:RR_and_m_k6} presents comparisons of the recovery rates 
of the majority voting neural networks.
The length of the sparse signal  is set to $n = 256$ and the sparseness parameter is set to $k = 6$.
From Fig.\ref{fig:RR_and_m_k6},
we can observe significant improvement in recovery performance 
compared with the performance of the single neural network.
A single feedforward neural network discussed in the previous section 
provides recovery rate around $50\%$ at $m=120$.
On the other hand, the majority voting neural networks with 
3 component nets achieves  $95\%$ at $m=120$.
The majority nets with 5 components shows further improvement to $99\%$ at $m = 120$.
This result implies that the soft majority voting process introduced in this section 
is effective to improve the reconstruction performance.
Another implication obtained from this result is that independently trained nets tend to have 
different estimation error patterns. This property explains  the improvement in recovery rate observed in this experiment.
We can see that there is still gap between the curves of the IP-based sparse recovery and 
the majority voting nets with $S=5$. 
The gap might be considered as the price we need to pay 
for obtaining reduction in required computing resources.

\begin{figure}[tbp]
        \begin{center}
          \includegraphics[clip, width=8.4cm]{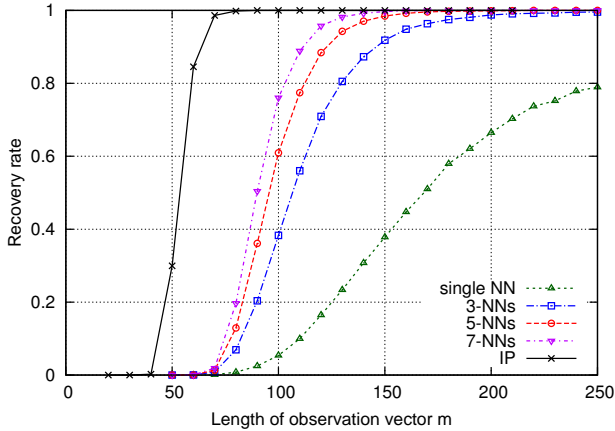}
          \caption{Recovery rate of majority voting neural network.  ($n=256, k=8)$}
          \label{fig:RR_and_m_k8}
        \end{center}
\end{figure}

Fig.\ref{fig:RR_and_m_k8} also shows the recovery rates when $k = 8$. The length of the original signal is $n = 256$.
In Fig.\ref{fig:RR_and_m_k8},
we can see the same tendency that has been observed in the previous experimental result in Fig.\ref{fig:RR_and_m_k6}.
The performance of sparse recovery tends to increase as the number of component networks grows.
At $m=120$, the recovery rate is improved from  $20\%$ (with the  single network) to $95\%$ (with 
$7$ component networks).

Table \ref{comptime} presents statistics on computation time required for sparse recovery of $10000$ instances 
for $n= 256, k =8$. 
It can be seen that sparse recovery algorithms based on neural networks runs order of 
several magnitude faster than the IP sparse recovery method. Of course, computation time 
depends on the computing environment and implementation but the result can be seen as an 
evidence that supports our claim that the proposed network structure is advantageous to reduce 
required computing resources.
\begin{table}[tbp]
\begin{center}
  \caption{Computation time (in second) for sparse recovery of $10000$ instances ($n= 256, k = 6$).  
  The parameter $S$ represents the number of component networks}
  \label{comptime}
  \begin{tabular}{r|rrrr} \hline \hline
    $m$ & IP & $S=1$ & $S=3$ &$S =5$  \\ \hline
    $60$ & $1.5 \times 10^4$ & $1.4 \times 10^{-1}$ & $9.7 \times 10^{-1}$ & $1.5 \times 10^0$ \\ \hline
    $90$ & $1.1 \times 10^4$ & $1.7 \times 10^{-1}$ & $8.7 \times 10^{-1}$ & $1.5 \times 10^0$ \\ \hline
    $120$ & $5.2 \times 10^3$ & $1.6 \times 10^{-1}$ & $8.9 \times 10^{-1}$ & $1.5 \times 10^0$ \\ \hline
    $150$ & $3.2 \times 10^3$ & $1.5 \times 10^{-1}$ & $9.7 \times 10^{-1}$ & $1.8 \times 10^0$ \\ \hline
    $180$ & $3.1 \times 10^3$ & $1.5 \times 10^{-1}$ & $9.4 \times 10^{-1}$ & $1.7 \times 10^0$ \\ \hline
  \end{tabular}
\end{center}
{\footnotesize The processor is Intel Core i7-3770K CPU(3.50GHz,  8-cores) and the memory size 
is 7.5 Gbytes.}
\end{table}

%
%

\section{Concluding summary}

In this paper,
we proposed sparse signal recovery schemes based on neural networks for binary compressed sensing.
Our empirical study shows a choice of the loss function used for training neural networks is 
of prime importance to achieve excellent reconstruction performance. We found a loss function 
suitable for this purpose, which includes a cross entropy like term and an $L_1$ regularized term. 
The majority voting neural network proposed in this paper 
is composed from several independently trained feedforward 
neural networks.  From the experimental results, we observed that  the majority voting neural network
achieves excellent recovery performance, which is approaching the optimal IP-based performance 
as the number of component nets grows.
The simple architecture of the majority voting neural network would be beneficial for 
both software and hardware implementation.
It can be expected that high speed sparse signal recovery circuits based on the neural networks 
produce novel applications in wireless communications such as multiuser detection in multiple access channels.

\section*{Acknowledgement}

The present study was supported by Grant-in-Aid for Scientific Research (B) (grant number 16H02878) from JSPS.
We used the optimization problem solver CPLEX Optimizer and the distributed numerical computation framework Tensorflow
in this work.
We gratefully acknowledge IBM Academic Initiative and Google.


\begin{thebibliography}{99}
\bibitem{onoff} A. K. Fletcher, S. Rangan and V. K. Goyal,
``On-off random access channels: a compressed sensing framework,''
arXiv:0903.1022, 2009.
\bibitem{RFID} M. Kaneko, W. Hu, K. Hayashi and H. Sakai,
``Compressed sensing-based tag identification protocol for a passive RFID system,''
IEEE Commun. Lett., vol. 18, no. 11, pp. 2023--2026,  2014.
\bibitem{1bit_comp} P. Boufounos and R. Baraniuk,
``1-bit compressive sensing,''
42nd Annual Conference on Information Sciences and Systems (CISS), pp. 16--21, 2008.
\bibitem{IP} Y. Plan and R. Vershynin, 
``One-bit compressed sensing by linear programming,''
Communications on Pure and Applied Mathematics 66.8, pp. 1275--1297, 2013. 
\bibitem{greedy}  P. Boufounos,
``Greedy sparse signal reconstruction from sign measurements''
Asilomar Conf. on Signals Systems and Comput., pp. 1305--1309, 2009.
\bibitem{BIHT} L. Jacques, J. Laska, P. Boufounos, and R. Baraniuk, 
``Robust 1-bit compressive sensing via binary stable embeddings of sparse vectors,''
IEEE Transactions on Information Theory 59.4,  2082--2102, 2013.
\bibitem{RSS} J. Laska, Z. Wen, W. Yin and R. Baraniuk, 
``Trust but verify: fast and accurate signal recovery from 1-bit compressive measurements,'' 
IEEE Trans. Signal Process., vol. 59, no. 11, pp. 5289--5301, 2011.
\bibitem{IHT}  T. Blumensath and M. Davies, 
``Iterative hard thresholding for compressive sensing,''
Appl. Comput. Harmon. Anal., vol. 27, no. 3, pp. 265--274, 2009.

\bibitem{deep_net} L. Deng, G. Hinton and B. Kingsbury,
``New types of deep neural network learning for speech recognition and related applications: an overview,''
International Conference on Acoustics, Speech, and Signal Processing (ICASSP),
pp. 8599--8603, 2013.

\bibitem{very_deep} K. Simonyan and A. Zisserman, 
``Very deep convolutional networks for large-scale image recognition,'' 
arXiv:1409.1556, 2014.
\bibitem{adam} D. P. Kingma and J. L. Ba,
``Adam: a method for stochastic optimization,''
arXiv:1412.6980, 2014.
\bibitem{tens}
``TensorFlow: large-scale machine learning on heterogeneous systems,''
\url{http://tensorflow.org/}
2015. Software available from tensorflow.org.
\bibitem{FPGA} G. Orchard, J. G. Vogelstein and R. Etienne-Cummings,
``Fast neuromimetic object recognition using FPGA outperforms GPU implementation,''
IEEE Trans. Neural Networks Learn. Syst., vol. 24, no. 8, pp. 1239--1252, 2013.
\bibitem{majority} S. B. Cho and J. H. Kim,
``Combining multiple neural networks by fuzzy integral for robust classigication,''
IEEE Trans. Systems Man and Cybernetics, vol. 25, no. 2, pp. 380--384, 1995.
\end{thebibliography}
\end{document}